%% file: RelayLowSNR-ISITLongArXivProof-20100428.tex
\documentclass[10pt,conference]{./sty/IEEEtran}

\usepackage{graphicx,graphics,epsfig,psfrag,subfigure}
\usepackage[usenames]{color}
\usepackage{amsmath,dsfont,amssymb}
\usepackage{cite,url,balance}
\usepackage{array,multirow,fancybox}
\usepackage[active]{srcltx}

\input{sty/macros}

\begin{document}

\title{On the Non-Coherent Wideband Multipath Fading Relay Channel}

\author{
\authorblockN{Nadia Fawaz, %\authorrefmark{1},
Muriel M{\'e}dard} %\authorrefmark{1}}
\authorblockA{%\authorrefmark{1}
Research Laboratory of Electronics,
Massachusetts Institute of Technology,
Cambridge, MA, USA%, \\
}
\authorblockA{Email: \{nfawaz,medard\}@mit.edu} %\authorrefmark{1}\{nfawaz,medard\}@mit.edu}
}

% make the title area
\maketitle

\begin{abstract}

We investigate the multipath fading relay channel  in the limit of a large bandwidth, and in the non-coherent setting, where the channel state is unknown to all terminals, including the relay and the destination. We propose a hypergraph model of the wideband multipath fading relay channel, and show that its min-cut is achieved by a non-coherent peaky frequency binning scheme. The so-obtained lower bound on the capacity of the wideband multipath fading relay channel turns out to coincide with the block-Markov lower bound on the capacity of the wideband frequency-division Gaussian (FD-AWGN) relay channel. In certain cases, this achievable rate also meets the cut-set upper-bound, and thus reaches the capacity of the non-coherent wideband multipath fading relay channel.

\end{abstract}

%\begin{keywords}
%
%\end{keywords}

\section{Introduction}\label{sec:Introduction}

The general relay channel is among the smallest building blocks of communication networks, yet its capacity is still an open problem. Bounds on the capacity of the general relay channel, and the capacity of some particular classes of relay channels, have been derived in the past \cite{Cover-ElGamal-1979}. In particular in \cite{ElGamal-Mohseni-ITtrans2006}, the expression of the cut-set upper bound from \cite{Cover-ElGamal-1979}, and the generalized block-Markov lower bound were derived for the case of the frequency-division additive white Gaussian noise (FD-AWGN) relay channel, where the source and the relay transmit in different bands. However, despite a plethora of recent works proposing cooperative strategies for wireless relaying networks and studying their performance in the high SNR regime, the capacity of the multipath fading relay channel remains unknown. Specifically, few works \cite{Avestimehr-Tse-ITtrans2007} analyzed the fading relay channel in the low SNR regime.

This paper focuses on analyzing the multipath fading relay channel in the non-coherent setting, where neither the source, nor the relay, nor the destination have channel state information (CSI), and in the wideband regime, alternatively named low SNR regime. Indeed, in the wideband regime, power is shared among a large number of degrees of freedom, making the SNR per degree of freedom low. Thus the wideband regime is power limited, but not interference limited on the contrary to the high SNR regime.
In the wideband regime, the capacity of the point-to-point AWGN channel  \cite{Shannon-1949} and the capacity of the point-to-point non-coherent multipath fading channel \cite{Kennedy-1969}
were shown to be both equal to the received SNR: $C_{Fading}=C_{AWGN}= \frac{P}{N_0} = \lim_{W \rightarrow \infty} W \log(1+\frac{P}{W N_0})$.
Moreover, in the wideband limit of fading channels, spread-spectrum signals were shown to achieve poor performance, whereas peaky signals in time and frequency, such as low duty-cycle FSK, along with non-coherent detection, were shown to be capacity optimal \cite{Telatar-Tse-2000}. The capacity of the point-to-point multiple input multiple output (MIMO) channel in the wideband limit was addressed in \cite{Zheng-Tse-2002}. In particular, for the SIMO channel with two receive antennas with respective gains $1$ and $a^2$, the capacity is $C_{SIMO}= (1+a^2) \frac{P}{N_0}$.
Results on multiple user channels in the wideband limit include, the capacity region of the AWGN Broadcast Channel (BC) \cite{McEliece-Swanson-1987}, for which time-sharing was shown to be optimal, and the capacity region of the AWGN Multiple Access Channel (MAC) \cite{Gallager-ITtrans1985}, for which FDMA allows all sources to achieve their point-to-point interference-free capacity to the destination.

\begin{figure*}[tp]
\begin{center}
\psfrag{S}[cc][cc]{{\small S}}
\psfrag{R}[cc][cc]{{\small R}}
\psfrag{D}[cc][cc]{{\small D}}
\psfrag{x_S}{$x_S$}
\psfrag{x_R}{$x_R$}
\psfrag{P_S}{{\small $P_S$}}
\psfrag{P_R}[cc][cc]{{\small $P_R=\gamma P_S$}}
\psfrag{y_{RS}}{$y_{RS}$}
\psfrag{y_{DS}}{$y_{DS}$}
\psfrag{y_{DR}}{$y_{DR}$}
\psfrag{h_{RS}}[br][bc]{$h_{RS}$}
\psfrag{h_{DS}}{$h_{DS}$}
\psfrag{h_{DR}}{$h_{DR}$}
\psfrag{N0}{{\small $N_0$}}
\subfigure[Multipath fading relay channel]{
\includegraphics[width=0.65\columnwidth]{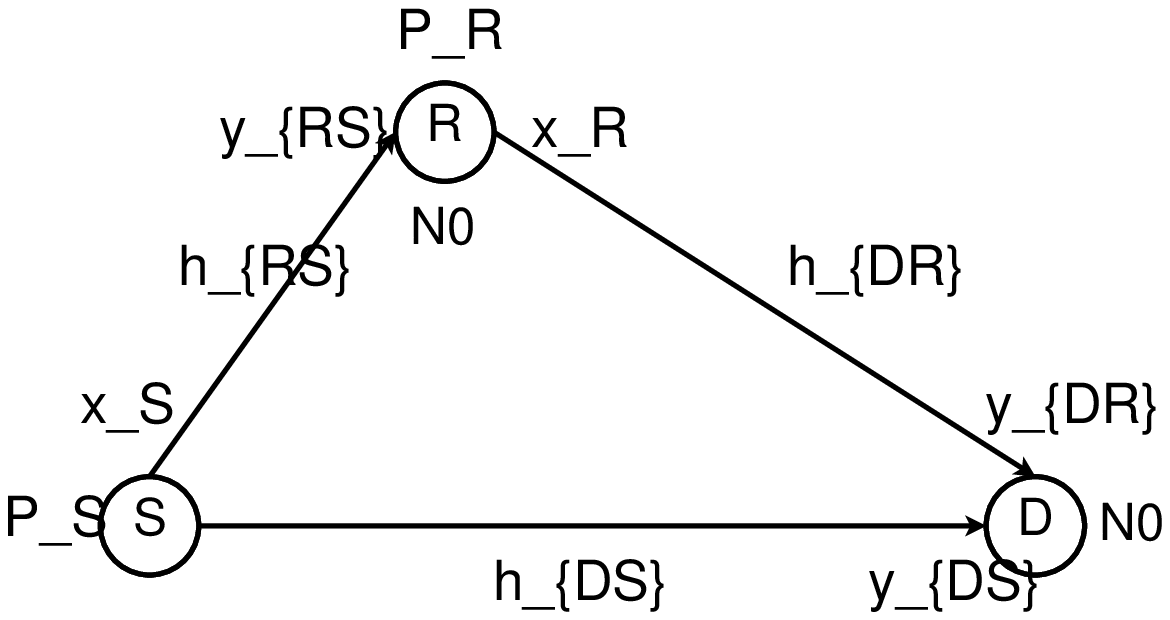}
%\caption{Fading relay channel}
\label{fig:SRD}
}
\qquad \qquad
\psfrag{S}[cc][cc]{{\small S}}
\psfrag{R}[cc][cc]{{\small R}}
\psfrag{D}[cc][cc]{{\small D}}
\psfrag{m1}[cc][cc]{{\small $m_1$}}
\psfrag{m2}[cc][cc]{{\small $m_2$}}
\psfrag{m3}[cc][cc]{{\small $\hat{m}_1$}}
\psfrag{P_S}{{\small }}
\psfrag{P_R}{{\small }}
\psfrag{b}{{\small \green{$\gamma b^2 \frac{P_S}{N_0}$}}}
\psfrag{(|a|^2-1)^+}[cc][cc]{{\small \red{$(a^2-1)^+ \frac{P_S}{N_0}$}}}
\psfrag{(1-|a|^2)^+}{{\small $\frac{P_S}{N_0} \ind_{[0,1]}(a^2)  $}}
\psfrag{min{1,|a|^2}}{{\small \blue{$\frac{P_S}{N_0} \ind_{]1,+\infty[}(a^2) $}}}
\psfrag{N0}{{\small }}
\psfrag{X}{{\small }}
\psfrag{x_R}{{\small }}
\psfrag{Y1}{{\small }}
\psfrag{y_{DS}}{{\small }}
\psfrag{y_{DR}}{{\small }}
%\psfrag{X}{{\small $X_S$}}
%\psfrag{x_R}{{\small $X_R$}}
%\psfrag{Y1}{{\small $Y_{RS}$}}
%\psfrag{y_{DS}}{{\small $Y_{DS}$}}
%\psfrag{y_{DR}}{{\small $Y_{DR}$}}
%\psfrag{h_{DS}}{{\small $1$}}
%\psfrag{h_{DR}}{{\small $b$}}
\subfigure[Hypergraph model of the wideband multipath fading relay channel. \emph{ \scriptsize For the sake of clearness, $(1-2T_d/T_c)$ is not displayed in edge capacities.}]{
\includegraphics[width=0.65\columnwidth]{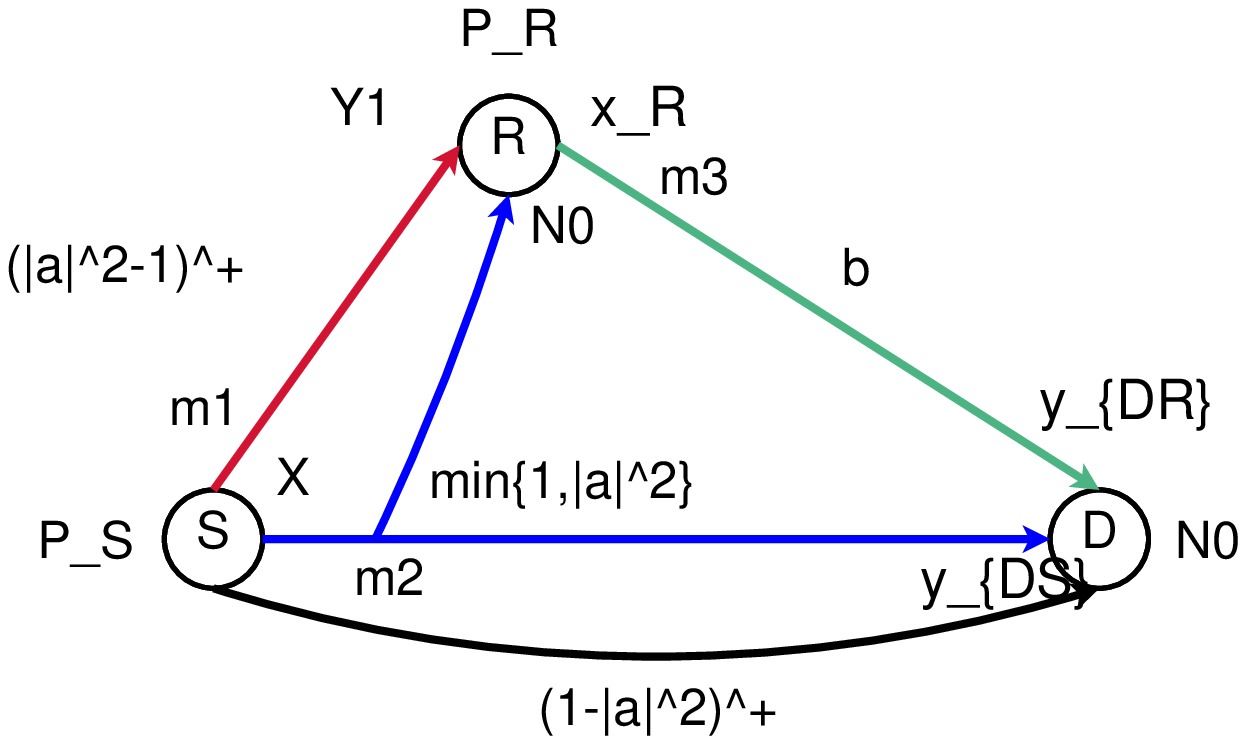}
%\caption{Hypergraph model of the wideband multipath fading relay channel}
\label{fig:Hypergraph}
}
\label{fig:FadingRelayCh}
\vspace{-0.15cm}
\caption[Fading relay channel]{Wideband fading relay channel}
\end{center}
\vspace{-0.8cm}
\end{figure*}

Some observations can be drawn from previous works on point-to-point and multiple user channels in the wideband regime: the capacity in the multipath fading case is the same as in the AWGN case, it can be reached in a non-coherent setting, and interference is not an issue.
Coming back to the non-coherent multipath fading relay channel in the wideband regime, two questions naturally arise
\begin{itemize}
\item Can the FD-AWGN lower bound \cite{ElGamal-Mohseni-ITtrans2006} be achieved in the non-coherent multipath fading case?

\item Can the cut-set upper-bound \cite{Cover-ElGamal-1979} be reached?
\end{itemize}
Note that in the wideband regime, considering the FD channel is relevant and meets the relay half-duplex constraint. This paper addresses these questions through three main contributions:
\begin{enumerate}
\item  A hypergraph model of the wideband multipath fading relay channel is proposed.

\item The hypergraph min-cut is shown to be achieved in the non-coherent wideband multipath fading relay channel by a peaky frequency-binning scheme.

\item The hypergraph min-cut is shown to coincide with the generalized block-Markov lower bound on the capacity of the wideband FD-AWGN relay channel, and in certain channel configurations with the cut-set upper-bound, in which case it is equal to capacity.

\end{enumerate}

% Outline
%------------------%

The rest of the paper is organized as follows.  In Section \ref{sec:SysMod}, the hypergraph model of the wideband multipath fading relay channel is described, and the achievable hypergraph min-cut is compared with bounds on the capacity of the wideband FD-AWGN relay channel. The non-coherent scheme achieving the hypegraph min-cut is described in Section \ref{sec:RelScheme}, while its correspondence with the hypegraph model is detailed in Section \ref{sec:MinCutHypergraph}, leading to the concluding Section \ref{sec:Conclusion}. Finally, the proof of \emph{Theorem~\ref{th:AchRateFading}} is provided in Appendix~\ref{sec:ErrorProba}.

\section{System model and main results}\label{sec:SysMod}

%Notations
%------------------%
Notations: $\N$ and $\R$ denote the sets of non-negative
integers, and real numbers, respectively. Let $m\in \N$, the set of non-negative integers less or equal to $m$ is denoted $\N_m \triangleq \{0,\ldots,m\}$. The
subset  $[0,+\infty [$ of $\R$ is denoted by $\R^+$. Let $x \in \R$, $(x)^+ \triangleq \max \{ 0, x \}$. Let $S$ be a set, the indicator function is defined by $\ind_S(x)=1$ if $x\in S$, $\ind_S(x)=0$ if $x\notin S$. $\Pr\{A\}$ is the probability of event A, $\E[\cdot]$ is the statistical expectation operator, and $X$ is $\mathcal{CN}(\mu,\sigma^2)$ means that $X$ is a circularly symmetric complex Gaussian random variable with mean $\mu$ and variance $\sigma^2$.

\subsection{Wideband multipath fading relay channel}

Consider the three-node network in Figure \ref{fig:SRD}, where the source S, the relay R and the destination D are equipped with a single antenna.
Source and relay are assumed to have average power constraints in the time-continuous channel model of $P_S$ and $P_R=\gamma P_S$ Joules/s respectively.
We assume that S, R and D have no channel state information (CSI), thus the multipath channel is considered in the non-coherent regime.
In order to respect the half-duplex constraint at the relay, we assume that S and R transmit in two different frequency bands of respective width $W_S$ and $W_R$. During each temporal block of duration $T$,
S transmits a new codeword which R and D receive in the first frequency band; R performs some transformation on the signal received from S in the previous block and relays it to D in the second frequency band; D decodes a new codeword by processing the signals it received from S and R.

As in \cite{Telatar-Tse-2000} the continuous-time multipath fading channel between transmitter $u\in \{S,R\}$ and receiver $v \in\{R,D\}$ is represented by the impulse response
%$ h_{vu}(t)=\sum_{l=1}^{L_{vu}} a_{vu,l}(t) \delta(t-d_{vu,l}(t))$,
\begin{equation}\label{eq:cir}
h_{vu}(t)=\sum_{l=1}^{L_{vu}} a_{vu,l}(t) \delta(t-d_{vu,l}(t)),
\end{equation}
where $L_{vu}$ is the number of paths, and $a_{vu,l}(t)$ and $d_{vu,l}(t)$ are the gain and delay of path $l$ at time $t$.
For the sake of simplicity, we assume that all channels $h_{vu}$, $u\in \{S,R\}$, $v \in\{R,D\}$ have similar coherence-time $T_c$ and delay-spread $T_d$.
Moreover we consider a block-fading model where the processes $\{a_{vu,l}(t)\}$ and $\{d_{vu,l}(t)\}$ have constant values $\{a_{vu,l}(nT_c)\}$ and $\{d_{vu,l}(nT_c)\}$ over intervals $[nT_c,(n+1)T_c[$. Furthermore, the processes $\{a_{vu,l}(nT_c)\}$ and $\{d_{vu,l}(nT_c)\}$ are assumed to be independent, stationary and ergodic. Finally, let $a,b \in \Rplus$, we assume a non-symmetric network, with stationary total channel gains $\sum_{l=1}^{L_{DS}} \E[|a_{DS,l}(0)|^2]=1$, $ \sum_{l=1}^{L_{RS}} \E[|a_{RS,l}(0)|^2]=a^2$, $ \sum_{l=1}^{L_{DR}} \E[|a_{DR,l}(0)|^2]=b^2$.

A signal $x_u(t)$ transmitted in channel $h_{vu}(t)$ leads a received signal $y_{vu}(t)=\sum_{l=1}^{L_{vu}} a_{vu,l}(t) x_u(t-d_{vu,l}(t)) + z_v(t)$,
%given by
%\begin{equation}\label{eq:yvu}
%y_{vu}(t)=\sum_{l=1}^{L_{vu}} a_{vu,l}(t) x_u(t-d_{vu,l}(t)) + z_v(t),
%\end{equation}
where $z_v(t)$ is a white Gaussian noise process with power spectral density $N_0/2$.

As the band grows large, the capacity of the point-to-point non-coherent wideband multipath fading channel is  equal to the received SNR \cite{Telatar-Tse-2000}. Thus, the capacities of the point-to-point wideband channels between the source and the destination, the source and the relay, and the relay and the destination are respectively $C_{DS}=\frac{P_S}{N_0}$, $C_{RS}=a^2 \frac{P_S}{N_0}$, and $C_{DR}=b^2 \frac{P_R}{N_0}$.

\subsection{Hypergraph model and main results}\label{sec:AchRate}

In this section, we introduce a hypergraph model of the wideband multipath fading relay channel, and gather our main results in \emph{Theorem \ref{th:AchRateFading}}. More precisely, we show that the hypergraph min-cut is achieved by a non-coherent relaying scheme based on peaky signals, which is described in details in Section \ref{sec:RelScheme}, and we compare the hypergraph min-cut with bounds on the capacity of the FD-AWGN relay channel.

The proposed hypergraph model of the wideband relay channel is depicted in Figure \ref{fig:Hypergraph}. A hyperedge connects a transmitting node to several receiving nodes. A message transmitted over a hyperedge at a rate below its capacity can be decoded reliably by all the receiving nodes. Messages transmitted over disjoint hyperedges are independent. This hypergraph model of the relay channel is motivated by the broadcast nature of the wireless link: when a source transmits a signal over the wireless link, several receiving nodes can overhear the signal and extract some of the information transmitted by the source. The hypergraph model allows to clarify the correlation between the pieces of information decoded at different receiving nodes, by breaking the wireless link from a transmitting node into a set of hyperedges carrying independent messages.
In Figure \ref{fig:Hypergraph}, the blue hyperedge represents a reliable channel from the source to both the relay and the destination with capacity $\frac{P_S}{N_0} \ind_{]1,+\infty[}(a^2) $, while the red and black edges represent extra reliable channels to the relay only with capacity $(a^2-1)^+\frac{P_S}{N_0}$, and to the destination only with capacity $\frac{P_S}{N_0} \ind_{[0,1]}(a^2) $, respectively. Note that the black channel cannot coexist simultaneously with the red and blue channels. Finally, the green edge represents a reliable channel from the relay to the destination with capacity $\gamma b^2 \frac{P_S}{N_0}$.

\begin{theorem}\label{th:AchRateFading}
Consider the non-coherent wideband multipath fading relay channel, described in Section \ref{sec:SysMod}. When the system bandwidth $W_S+W_R$ grows large,
\begin{enumerate}
\item a lower bound on the capacity is provided by the min-cut on the hypergraph model
\begin{equation}\label{eq:LBfading}
\hspace{-0.4cm} R = \min\left\{ \max\{1,a^2\} , (1+b^2 \gamma) \right\}\frac{P_S}{N_0} \left(1-2\frac{T_d}{T_c}\right).
\end{equation}
\item this achievable rate (\ref{eq:LBfading}) is equal to the wideband limit of the generalized block Markov lower bound of the FD-AWGN channel \cite{ElGamal-Mohseni-ITtrans2006} with the same received SNRs in the point-to-point source-destination, source-relay, and relay-destination channels when the channel is underspread ($T_d \ll T_c$).
\item in the case where $a^2 \geq 1+b^2 \gamma$ and $T_d \ll T_c$, this achievable rate (\ref{eq:LBfading}) is equal to the FD-AWGN cut-set upper-bound $(1+b^2 \gamma) \frac{P_S}{N_0}$, and it is therefore the capacity of the non-coherent wideband multipath fading channel.
\end{enumerate}
\end{theorem}
The proof of $1)$ in \emph{Theorem \ref{th:AchRateFading}} is provided in Appendix \ref{sec:ErrorProba}. We now address 2) and 3).
The cut-set upper bound, and the generalized block-Markov lower bound on the capacity of the FD-AWGN relay channel were derived in \cite{ElGamal-Mohseni-ITtrans2006}. When the system bandwidth grows large, the cut-set upper bound converges to
\begin{equation}\label{eq:FDAWGNcutSetB}
 C_{FD-AWGN} \leq \min\left\{(1+a^2), (1+\gamma b^2)\right\}\frac{P_S}{N_0},
\end{equation}
and the generalized block-Markov lower bound converges to
\begin{equation}\label{eq:FDAWGNlowerB}
C_{FD-AWGN} \geq \min\left\{\max\{1,a^2\},(1+\gamma b^2) \right\}\frac{P_S}{N_0}.
\end{equation}
Comparing (\ref{eq:LBfading}) and (\ref{eq:FDAWGNlowerB}) shows that the lower bounds on the capacity of the non-coherent multipath fading relay channel and the FD-AWGN channel coincide in the wideband limit when the channel is underspread ($T_d \ll T_c$). This justifies 2) in \emph{Theorem \ref{th:AchRateFading}}, and shows that the hypergraph model is also valid in the FD-AWGN case.

In the case where $a^2 \geq 1+b^2 \gamma$ and the channel is underspread ($T_d \ll T_c$), the bounds (\ref{eq:LBfading}), (\ref{eq:FDAWGNlowerB}) and (\ref{eq:FDAWGNcutSetB}) coincide. The capacity of the multipath fading relay channel with infinite bandwidth cannot exceed the cut-set upper bound of the infinite bandwidth AWGN relay channel. We can then conclude  that $(1+b^2 \gamma) \frac{P_S}{N_0}$ is the capacity of the non-coherent wideband multipath fading channel in that case, as stated in 3) in \emph{Theorem \ref{th:AchRateFading}}.

The multipath fading achievable rate (\ref{eq:LBfading}) and the FD-AWGN cut-set upper-bound  (\ref{eq:FDAWGNcutSetB}) are plotted in Figure \ref{fig:BoundsMultipath} in blue and red respectively, in the case where $1<\gamma b^2$.

\section{Relaying scheme achieving the min-cut}\label{sec:RelScheme}

In this section, we describe the non-coherent scheme which achieves the hypergraph min-cut in the multipath fading case.

\subsection{Peaky Signaling at Source}

\begin{figure}[tp]
%\begin{center}
\begin{flushright}
\psfrag{m=m1*M1+m2}{{\footnotesize $m=m_1M_R+m_2$}}
\psfrag{m1}[bc][br]{\red{\footnotesize $m_1$}}
\psfrag{m2}[bc][br]{\blue{\footnotesize $m_2$}}
\psfrag{msg m}[br][br]{{\scriptsize $m\in\N_{M_S-1}$}} %\{0,\ldots,M_S-1\}
\psfrag{index m1}[br][br]{\red{\scriptsize   $ m_1\in \N_{M_R-1} $}} %\{0,\ldots,M_R-1\}
\psfrag{index m2}[br][br]{\blue{\scriptsize $ m_2\in \N_{M_D-1} $}} %\{0,\ldots,M_D-1\}
\psfrag{0r}[bc][bc]{\red{ \scriptsize $0$}}
\psfrag{1r}[bc][bc]{\red{ \scriptsize $1$}}
\psfrag{M1-2}[bc][bc]{\red{ \scriptsize  $M_R-2$}}
\psfrag{M1-1}[bc][bc]{\red{ \scriptsize $\:\: M_R-1$}}
\psfrag{0b}{\blue{ \scriptsize $0$}}
\psfrag{1b}{\blue{ \scriptsize $1$}}
\psfrag{M2-1}{\blue{ \scriptsize $M_D-1$}}
\includegraphics[width=0.9\columnwidth]{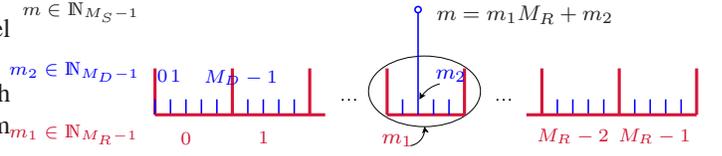}
%{\small {\red Fig. 6.} Frequency binning}
\vspace{-0.1cm}
\caption{{\normalsize Binning $m=(m_1,m_2)$}}
\label{fig:FreqBin}
\end{flushright}
%\end{center}
\vspace{-0.8cm}
\end{figure}

Let $M_S$, $M_R$ and $M_D$ be positive integers such that the codebook size at source is $M_S=M_R M_D$. Consider a couple of independent random integers $(m_1,m_2)$ such that $m_1\in\N_{M_R-1}\triangleq\{0,\ldots,M_R-1\}$, $m_2\in\N_{M_D-1}\triangleq\{0,\ldots,M_D-1\}$. Then the Euclidian division theorem ensures that there exists a unique source message $m\in\{0,\dots,M_S-1\}$ such that $m= m_1M_D+m_2$. The representation of $m$ as a couple $(m_1,m_2)$ has a binning interpretation. Indeed, the $M_S$ messages can be grouped into $M_R$ bins of $M_D=\frac{M_S}{M_R}$ messages. The integer $m_1$ represents the bin index of message $m$, while $m_2$ is the index of message $m$ within bin $m_1$, as illustrated in Figure \ref{fig:FreqBin}. For $m_1\in \N_{M_R-1}$, the $m_1$-th bin is denoted $\bin_{m_1}$ and contains the $M_D$ messages $\bin_{m_1}=\{m_1M_D,\ldots,m_1M_D+M_D-1\}$.

During the first block of the cooperative transmission scheme, the source transmits a message $m$ using the peaky-signaling scheme in \cite{Telatar-Tse-2000}, which was shown to achieve capacity in the wideband regime, and that we recall briefly in this section.
The transmission scheme is based on Frequency Shift Keying (FSK) and low-duty cycle, and is therefore peaky both in frequency and time. We denote by $\theta \in ]0,1]$ the duty-factor, representing the fraction of time during which the source actually transmits power. If the source transmits power during $T_s$, then the time separating two successive transmissions is $T_s/\theta$.
Using FSK the transmitted signal corresponding to the $m$-th message is given in the baseband by a sinusoid at frequency $f_m$ with power $P_S/\theta$
\begin{equation}
x_S(t)=
\left\{
    \begin{array}{ll}
        \sqrt{\frac{P_S}{\theta}} \exp(j 2 \pi f_m t) & , 0\leq t \leq T_s \\
        0 & , T_s\leq t \leq T_s/\theta,
    \end{array}
\right.
\end{equation}
where the transmission duration is chosen to be shorter than the coherence time $T_s \leq T_c$. Frequencies $f_m$ are taken to be integer multiples of $1/(T_s - 2 T_d)$, leading to a minimum  bandwidth $W_S = M_S / (T_s - 2 T_d)$ for a codebook size $M_S$.

During the interval $[T_d, T_s - T_d]$ the processes $\{a_{RS,l}(t)\}$ and $\{d_{RS,l}(t)\}$ are constant, thus the signal received by the relay, when message $m$ is sent, is given by:
\begin{equation}
\begin{split}
y_{RS}(t)& =\sum_{l=1}^{L_{RS}} a_{RS,l}  \sqrt{\frac{P_S}{\theta}} \exp(j 2 \pi f_m (t-d_{RS,l})) + z_R(t)\\
         %& = G_{RS} \sqrt{\frac{P_S}{\theta}} \exp(j 2 \pi f_m t) + z_R(t)\\
         & = G_{RS} x_S(t) + z_R(t), \nonumber
\end{split}
\end{equation}
where $G_{RS}=\sum_{l=1}^{L_{RS}} a_{RS,l} \exp(- j 2 \pi f_m d_{RS,l})$ is the complex gain of the source-relay channel during $[T_d,T_s-T_d]$.

Similarly, we define the complex gain for the source-destination channel $G_{DS}=\sum_{l=1}^{L_{DS}} a_{DS,l} \exp(- j 2 \pi f_m d_{DS,l})$ and the signal received by the destination during $[T_d, T_s - T_d]$, $y_{DS}(t)= G_{DS} x_S(t) + z_D(t).$
The source repeats the transmission of a symbol $N$ times over $N$ disjoint time intervals $\frac{T_s}{\theta}$ to obtain diversity. Both the relay and the destination receive the $N$ signals corresponding to message $m$ during the first temporal block of total duration $T=\frac{NT_s}{\theta}$.

To transmit a codeword carrying $\ln M_S$ nats of information, an average power $P_S$ is used, and the source rate is given by $R \triangleq \frac{\theta}{NT_s}\ln M_S$.
Note that the source rate can be written $R=R_1+R_2$, with  $R_1\triangleq \frac{\theta}{NT_s}\ln M_R$ and $R_2\triangleq \frac{\theta}{NT_s}\ln M_D$.

\subsection{Processing at Relay}\label{sec:ClipF}

Upon reception of the $N$ source signals, the relay first decodes the bin index $\hat{m}_1$, then it forwards $\hat{m}_1$ to the destination using peaky signaling. Note that if the number of bins was set to $M_R=1$ single bin, this would render the relay unused, and correspond to a direct transmission from S to D.

\smallskip

\noindent \textbf{Phase 1: Decoding $\hat{m}_1$ by correlating}

\noindent The relay correlates the $n^{th}$ received signal against each frequency $k\in\N_{M_S-1}$, forming the correlations
  \begin{eqnarray}\label{eq:corr}
  R_{RS,k}(n) & \triangleq & \!\!\! \frac{1}{\sqrt{N_0(T_s - 2 T_d)}}\int_{T_d}^{T_s - T_d} \!\!\!\!\!\!\!\!\! y_{RS}(t) exp(-j2 \pi f_k t) dt \nonumber \\
          & = & \!\!\! \delta_{km} \sqrt{\frac{P_S (T_s - 2 T_d)}{\theta N_0}} G_{RS}(n) + W_k(n),
  \end{eqnarray}
  where $G_{RS}(n)$ is the complex gain in interval $n$, $\{W_k(n)\}_n$ are i.i.d. circularly symmetric complex Gaussian random variables with unit-variance. By modeling assumption, $\{G_{RS}(n)\}_n$ are i.i.d. complex random variables. Assuming a large number of paths, $\{G_{RS}(n)\}_n$ can be modeled by i.i.d. circularly symmetric complex Gaussian random variables with $0$-mean and variance $a^2$. Then for each $k$, $\{R_{k}(n)\}$ are i.i.d. $\mathcal{CN}(0,\sigma_k^2)$ with variances
\vspace{-0.05cm}
  \begin{equation}\label{eq:sigmak}
  \sigma_k^2= 1+ \delta_{km} \frac{a^2 P_S (T_s - 2 T_d)}{\theta N_0} \mbox{ , }  k \in \N_{M_S-1}.
  \end{equation}
  Note that $\sigma_k^2=1$ for all $k \neq m$.
  The relay decoder builds the decision variables
\vspace{-0.25cm}
\begin{equation}
S_{RS,k}=\frac{1}{N}\sum_{n=1}^N |R_{RS,k}(n)|^2,
\end{equation}
which for all $k\neq m$ are i.i.d. These decision variables are compared with the threshold $A_R = 1+ (1-\epsilon) \frac{a^2 P_S (T_s - 2 T_d)}{\theta N_0}$, with $\epsilon \in ]0,1[$ to determine the set $\mathcal{S}_R$ of bins containing at least one frequency above threshold
\begin{equation}\label{eq:DecSetR}
\mathcal{S}_R \triangleq \{ k \in \N_{M_R-1} : \:\: \exists \: l \in \bin_k \mbox{ s.t. } S_{RS,l} \geq A_R \}.
\end{equation}
If $\mathcal{S}_R$ only contains a single bin $k$, the relay decodes $\hat{m}_1=k$, otherwise it declares an error.

\smallskip

\noindent \textbf{Phase 2: Forwarding the bin index  $\hat{m}_1$}

\noindent If the relay has not declared an error at the end of Phase 1, then it forwards the bin index $\hat{m}_1$ to the destination using peaky FSK in the second frequency band, with duty cycle $\theta$ and frequencies multiple of $1/(T_s - 2 T_d)$. Similarly to the source, the relay repeats $N$ times the transmission of $\hat{m}_1$ over disjoint intervals of duration $\frac{T_s}{\theta}$ for diversity. In the $n$-th interval, during the fraction $\theta$ of time where the relay signal is non-null, the signal is given by
%\begin{equation}\label{eq:xR}
$x_R(t)= \sqrt{\frac{P_R}{\theta}} \exp(j 2 \pi f_{\hat{m}_1} t)$.
%\end{equation}

During the interval $[(N+n-1)\frac{T_s}{\theta} + T_d, (N+n)\frac{T_s}{\theta}- T_d]$, of length $(T_s-2T_d)$, the signal received by the destination, corresponding to the $n$-th relay signal, can be written
%\begin{equation}
$y_{DR}(t)= G_{DR} x_R(t) + z_D(t)$,
%\end{equation}
where $G_{DR} = \sum_{l=1}^{L_{DR}} a_{DR,l} \exp(- j 2 \pi f_{\hat{m}_1} d_{DR,l})$ is the complex gain of the relay-destination channel.
To transmit a codeword carrying $\ln M_R$ nats of information, a minimum bandwidth $W_R = M_R / (T_s - 2 T_d)$ and an average power $P_R$ are used, and the relay rate is given by $R_1 = \frac{\theta}{NT_s}\ln M_R$.

\subsection{Decoding at Destination}\label{sec:Detection}

At the end of the second phase, the destination has received $2N$ signals corresponding to the same message $m$, half coming from the source, and half being the retransmissions from the relay. The destination first processes the signal from the relay to decode the bin index $m_1$, then the signal from the source to decode the remaining index $m_2$.

\smallskip

\noindent \textbf{Step 1: Decoding the bin index $\hat{\hat{m}}_1$}

\noindent Similarly to (\ref{eq:corr}), the destination correlates the $N$ signals from the relay against each of the $M_R$  frequencies in the second band, to form the correlations $R_{DR,k}(n)$, for $n \in \N_{N}$, and $k \in \N_{M_R-1}$, given by
\begin{equation}
\begin{split}
R_{DR,k}(n) &= \delta_{k \hat{m}_1} \sqrt{\frac{P_R (T_s - 2 T_d)}{\theta N_0}} G_{DR}(n) + W_{R,k}(n), \nonumber
\end{split}
\end{equation}
where $\{W_{R,k}(n)\}$ are i.i.d. $\mathcal{CN}(0,1)$.
Assuming a large number of paths, $\{G_{DR}(n)\}_n$ are modeled by i.i.d. $\mathcal{CN}(0,b^2)$ random variables. Then, for each $k\in \N_{M_R-1}$, the variables $\{R_{k}(n)\}_n$ are i.i.d. $\mathcal{CN}(0,\sigma_{R,k}^2)$ with variances
$\sigma_{R,k}^2= 1+ \delta_{k\hat{m}_1} \frac{b^2 P_R (T_s - 2 T_d)}{\theta N_0}$.
The destination compares the decision variables $S_{DR,k} = \frac{1}{N}\sum_{n=1}^N |R_{DR,k}(n)|^2$ with the threshold $B_R = 1+ (1-\epsilon_1) \frac{b^2 P_R (T_s - 2 T_d)}{\theta N_0}$ and builds the set
\begin{equation}\label{eq:DecSet1}
\mathcal{S}_1=\{ k \in \N_{M_R-1} \: : \:\:  S_{DR,k} \geq B_R \}.
\end{equation}
If $|\mathcal{S}_{1}|=1$, the destination decodes $\hat{\hat{m}}_1$, otherwise it declares an error.

\smallskip

\noindent \textbf{Step 2: Decoding the index $\hat{m}_2$}

\noindent If the destination has not declared an error at the end of Step~1, it can proceed with the decoding by processing the signal it received from the source in the previous block. The destination uses $\hat{\hat{m}}_1$ to locate the bin of $M_D$ frequencies containing the source message $m$ in the signal $y_{DS}$.
The destination correlates the $N$ messages it received from the source against the $M_D$ frequencies in $\bin_{\hat{\hat{m}}_1}=\{\hat{\hat{m}}_1 M_D,\ldots,\hat{\hat{m}}_1 M_D+M_D-1\}$ to form the correlations $R_{DS,l}(n)$, for $n \in \N_{N}$, and $l \in \bin_{\hat{\hat{m}}_1}$
\begin{equation}
R_{DS,l}(n)= \delta_{lm} \sqrt{\frac{P_S (T_s - 2 T_d)}{\theta N_0}} G_{DS}(n) + W_{S,l}(n),
\end{equation}
where $\{W_{S,l}(n)\}_n$  are $\mathcal{CN}(0,1)$, and for each $l$, the variables $\{R_{DS,l}(n)\}$ are i.i.d. $\mathcal{CN}(0,\sigma_{S,l}^2)$ with variance $\sigma_{S,l}^2= 1+ \delta_{lm} \frac{P_S (T_s - 2 T_d)}{\theta N_0} \mbox{ , }  l \in \bin_{\hat{\hat{m}}_1}$.
%\begin{equation}
%\sigma_{S,l}^2= 1+ \delta_{lm} \frac{P_S (T_s - 2 T_d)}{\theta N_0} \mbox{ , }  l \in \bin_{\hat{\hat{m}}_1}.
%\end{equation}
It should be pointed out that the relayed signal allows the destination to reduce the dimension of the space in which it looks for the source message $m$. More precisely, the relayed message allows the destination to reduce the number of noisy frequencies, to which it needs to compare the signal $y_{DS}$, from $M_S=M_RM_D$ to $M_D$. This observation is critical in the wideband regime where performance is mainly impaired by noise.
For $l \in \bin_{\hat{\hat{m}}_1}$, the destination builds the decision variables $S_{DS,l} = \frac{1}{N}\sum_{n=1}^N |R_{DS,l}(n)|^2$.
By comparing them with $B_S = 1+ (1-\epsilon_2) \frac{P_S (T_s - 2 T_d)}{\theta N_0}$, it builds the set
\begin{equation}\label{eq:DecSet2}
\mathcal{S}_2=\{ l \in \bin_{\hat{\hat{m}}_1} \: : \:\:  S_{DS,l} \geq B_S \}.
\end{equation}
If $|\mathcal{S}_{2}|=1$, the destination decodes $\hat{m}_2$, otherwise it declares an error.

If the destination decoder passes Steps 1 and 2  without declaring an error, the destination forms the final decoded message $\hat{m}=\hat{\hat{m}}_1 M_D + \hat{m}_2$.

\section{Hypergraph interpretation}\label{sec:MinCutHypergraph}

\begin{figure}[tp]
\begin{center}
\psfrag{(1+|b|^2)P/N0}[cc][cl]{{\small $(1+ \gamma b^2)\frac{P_S}{N_0}$}}
\psfrag{P/N0}{{\small $\frac{P_S}{N_0}$}}
\psfrag{(1+|a|^2)P/N0}[cc][cc][1][40]{{\small $(1+a^2)\frac{P_S}{N_0}$}}
\psfrag{|a|^2 P/N0}[cc][cc][1][40]{{\small $a^2\frac{P_S}{N_0}$}}
\psfrag{|a|^2}{{\footnotesize \begin{tabular}{l}
                   S-R channel \\
                   gain $a^2$ \\
                 \end{tabular}}}
\psfrag{1}[cc][cc]{{\small $1$}}
\psfrag{|b|^2}[cc][cc]{{\small $\gamma b^2$}}
\psfrag{1+|b|^2}[cc][cc]{{\small $1+\gamma b^2$}}
\psfrag{Rate [bps]}[cc][cl]{{\small Rate [bps]}}
\psfrag{SD rate}[cc][cc]{{\scriptsize \begin{tabular}{l}
                   S: tx at $C_{DS}$\\
                   R: $\varnothing$\\
                   D: Dec$(m)$\\
                 \end{tabular}}}
\psfrag{SR rate}[cc][cc]{{\scriptsize\begin{tabular}{l}
                   S: Tx at $C_{RS}$\\
                   R: Freq. Binning at\\ $C_{RS}-C_{DS}\leq C_{RD}$\\
                   D: Dec$(m_1->m_2)$\\
                 \end{tabular}}}
\psfrag{SR rate2}{}
\psfrag{SD+RD rate}{{\scriptsize\begin{tabular}{l}
                   S: Tx at $C_{DS}+C_{DR}$\\
                   R: Freq. Binning at\\ $C_{DR}\leq C_{RS}$\\
                   D: Dec$(m_1->m_2)$\\
                 \end{tabular}}}
%\psfrag{SD rate}{{\scriptsize \begin{tabular}{l}
%                   S: Direct tx at $C_{DS}$ \\
%                   R: $\varnothing$ \\
%                   D: Dec$(X)$\\
%                 \end{tabular}}}
%\psfrag{SR rate}{{\scriptsize\begin{tabular}{l}
%                   S: Tx at $C_{RS}$ \\
%                   R: DF at $C_{RS}$\\
%                   D: Dec$(X_1)$\\
%                 \end{tabular}}}
%\psfrag{SR rate2}{{\scriptsize\begin{tabular}{l}
%                   S: Tx at $C_{SR}$ \\
%                   R: Dec at $C_{SR}$\\
%                      Freq. Binning at $C_{RS}-C_{DS}\leq C_{RD}$\\
%                   D: Dec$(X_1->X)$\\
%                 \end{tabular}}}
%\psfrag{SD+RD rate}{{\scriptsize\begin{tabular}{l}
%                   S: Tx at $C_{DS}+C_{DR}$ \\
%                   R: Dec at $C_{DS}+C_{DR}\leq C_{RS}$\\
%                      Freq. Binning at $C_{DR}$\\
%                   D: Dec$(X_1->X)$\\
%                 \end{tabular}}}
\psfrag{CS UB}{{\scriptsize Cut-set UB}}
\psfrag{Freq. Binning LB}{{\scriptsize Frequency binning LB}}
\includegraphics[width=0.71\columnwidth]{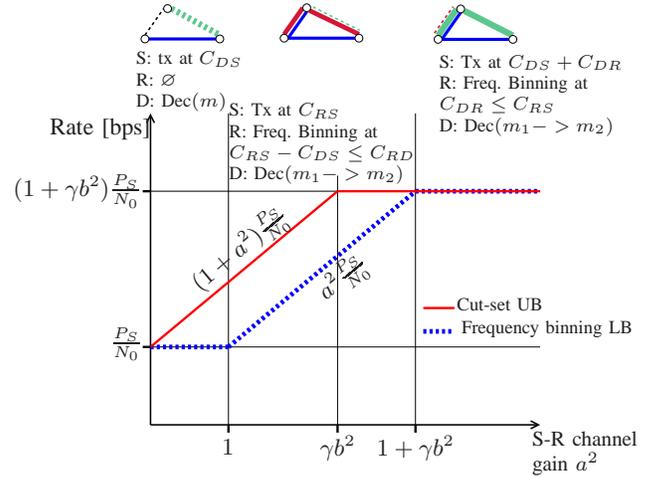}
%\begin{center}
%{\small {\r Fig. 7.} Bounds on capacity of non-coherent multipath fading channel }
%\end{center}
\vspace{-0.2cm}
\caption{{ Bounds on the capacity of the non-coherent wideband multipath fading relay channel}}
\label{fig:BoundsMultipath}
\end{center}
\vspace{-1cm}
\end{figure}

In this section, we give the correspondence between the min-cut achieving scheme in Section \ref{sec:RelScheme}, and the hypergraph model in Figure \ref{fig:Hypergraph}. The relaying scheme in Section \ref{sec:RelScheme} is a form of selective decode-and-forward, where the minimum amount of relayed information depends on the quality of the source-relay channel $C_{RS}=a^2 \frac{P_S}{N_0}$ with respect to the channels $C_{DS}=\frac{P_S}{N_0}$ and $C_{DR}=b^2\gamma \frac{P_S}{N_0}$. Indeed, the amount of information forwarded by the relay is parameterized by the value of $M_R$, relatively to $M_S=M_R M_D$. Three different regimes can be identified, as shown in Figure~\ref{fig:BoundsMultipath}.

\textbf{Regime $a^2 \leq 1$}: in this regime $C_{RS} \leq C_{DS}$, the source-destination channel is more reliable than the source-relay channel. The source transmits directly to the destination at capacity $C_{DS}=\frac{P_S}{N_0}$ without using the relay. This is equivalent to setting the number of bins to a single bin, $M_R=1$, containing all messages $M_D=M_S$. The achievable rate $R=C_{DS}=\frac{P_S}{N_0}$ is
given by the capacity of the black source-destination hyperedge.

\textbf{Regime $1 < a^2 \leq 1+b^2\gamma$}: in this regime $C_{DS} < C_{RS} \leq C_{DS}+C_{DR}$, the source-relay channel is stronger than the source-destination channel but weaker than the cut on the multiple-access (MA) side. The source transmits $m$ at rate $R=C_{RS}=a^2 \frac{P_S}{N_0}$, by splitting $m$ into submessages $m_2$ sent on the blue hyperedge at rate $R_2=C_{DS}=\frac{P_S}{N_0}$, and $m_1$ sent on the red hyperedge at rate $R_1=C_{RS}-C_{DS}= (a^2-1) \frac{P_S}{N_0}$. The relay decodes and reliably forwards the bin index $\hat{m}_1$ to the destination on the green hyperedge since $R_1\leq C_{DR}=b^2 \gamma \frac{P_S}{N_0}$. The destination will use the signals from relay and source to decode the remaining index $\hat{m}_2$. The number of bins is chosen $M_R \in ]1,M_S[$ such that $M_D$ matches the capacity of the source-destination channel, and $M_R$ can be handled by the source-relay and relay-destination channels. The achievable rate $R=C_{RS}=a^2 \frac{P_S}{N_0}$ is given by the sum of the capacities of the red and blue hyperedges.

%Note that, alternatively, the relay could also forward $\hat{m}_1$ at rate $C_{DR}=b^2 \gamma \frac{P_S}{N_0}$, and the destination would decode the remaining $\hat{m}_2$ at rate $C_{RS}-C_{DR} \leq C_{DS}$. In that case, the number of bins $ 1 < M_R < M_S$ would match the capacity of relay-destination channel.

%\smallskip

\textbf{Regime $1+b^2\gamma < a^2$}: in this regime $C_{DS}+C_{DR}<C_{RS}$, the source-relay channel is better than the multiple-access cut. The source transmits at a rate equal to the capacity of the MA cut $R=C_{DS}+C_{DR}=(1+b^2 \gamma) \frac{P_S}{N_0}$, by splitting $m$ into submessages $m_2$ sent on the blue hyperedge at rate $R_2=C_{DS}=\frac{P_S}{N_0}$, and $m_1$ sent on the red hyperedge at rate $R_1=C_{DR}= \gamma b^2 \frac{P_S}{N_0}\leq C_{RS}=a^2 \gamma \frac{P_S}{N_0}$.
The relay decodes and forwards the bin index $\hat{m}_1$ to the destination. The destination uses the signals from source and relay to decode the remaining $\hat{m}_2$. The number of bins $M_R \in ]1,M_S[$ matches the capacity of the relay-destination channel, and $M_D\in ]1,M_S[$ matches the capacity of the source-destination channel. The achievable rate $R=C_{DS}+C_{DR}=(1+b^2 \gamma) \frac{P_S}{N_0}$ is given by the sum of the capacities of the green and blue hyperedges.

In those three regimes, the achievable rate is given by the hypergraph min-cut. The relationship between the rate achieved by the peaky frequency binning scheme and the min-cut on the hypergraph appears as a simple tool to derive achievable rates, and the corresponding transmission schemes, in larger wideband relaying networks.

\section{Conclusion}\label{sec:Conclusion}

\vspace{-0.1cm}

We propose a hypergaph model of the relay channel in the wideband limit, and show that its min-cut can be achieved not only in the FD-AWGN case, but also in the non-coherent multipath fading case thanks to a relaying scheme combining peaky signals and binning.  In certain channel configurations, the so-obtained achievable rate also coincides with the cut-set upper-bound, and thus is equal to the capacity of the non-coherent wideband multipath fading channel.

In the remaining cases, where the rate achieved by the proposed scheme does not coincide with the cut-set upper bound, a question remains open: can the gap to the cut-set upper-bound be closed? If the capacity of the relay-destination channel was infinite, as in the SIMO channel, the cut $(1+a^2)\frac{P}{N_0}$ could be achieved, and the gap to the cut-set bound could be closed. However, because of the relay power constraint and the destination noise, the relay cannot make its received signal perfectly available to the destination as in the SIMO channel. This raises the question as to whether virtual MIMO gains can actually be achieved in the wideband regime.

\section*{Acknowledgments}

The work of Nadia Fawaz was supported by the French D{\'e}l{\'e}gation G{\'e}n{\'e}rale pour l'Armement (DGA), by IT-MANET project under subcontract $\sharp$ 18870740-37362-C issued by Stanford University and supported by the Defense Advanced Research Projects Agency (DARPA), and by  NeTS XORS  project supported by the National Science Foundation (NSF) under Grant No. CNS-0627021.

%%%%%%%%%%%%%%%%%%%%%%%%%%%%%%%%%%%%%%%%%%%%%%%%%%%%%%%%%%%
\appendices
%
%%%%%%%%%%%%%%%%%%%%%%%%%%%%%%%%%%%%%%%%%%%%%%%%%%%%%%%%%%%%%%%%%%%

\section{Proof of \emph{Theorem \ref{th:AchRateFading}}: Upper bound on the Probability of Error}\label{sec:ErrorProba}

In this appendix, an upper bound on the probability of error of the scheme described in Section \ref{sec:SysMod} is provided, and for all rates below (\ref{eq:LBfading}), this upper bound is shown to vanish to $0$ when the system bandwidth $W_S+W_R$ grows large.

If $a^2\leq 1$ the source transmits directly to the destination without using the relay by setting $M_R=1$, and the probability of error at the destination decays to $0$ as $N$ grows large if
\begin{equation}
\begin{split}
R \leq & C_{DS}  =\frac{P_S}{N_0} \left(1-2\frac{T_d}{T_c}\right) \\
 = & \min \left\{ \max\{1,a^2\}, 1+\gamma b^2 \right\} \: \frac{P_S}{N_0} \left(1-2\frac{T_d}{T_c}\right), \\
& \forall a^2 \leq 1 \mbox{ and } b^2 \geq 0.
\end{split}
\end{equation}

If $a^2 >1 $, the source transmits with the help of the relay, using the scheme described in Section \ref{sec:SysMod} with $M_R = \alpha M_S$ with $\alpha \in ]0,1]$. Assuming that message $m$ was sent by the source, the probability of error $\Pr\{ \hat{m} \neq m\}$ at the destination decoder can be written
\begin{equation}\label{eq:Pe}
\begin{split}
P_e=\Pr\{ \hat{m} \neq m\} =
& \Pr\{\hat{m} \neq m | \hat{\hat{m}}_1 = m_1 \} \Pr\{ \hat{\hat{m}}_1 = m_1\}\\
& + \Pr\{\hat{m} \neq m | \hat{\hat{m}}_1 \neq m_1 \} \Pr\{ \hat{\hat{m}}_1 \neq m_1\}\\
\leq & \Pr\{\hat{m} \neq m | \hat{\hat{m}}_1 = m_1 \} + \Pr\{ \hat{\hat{m}}_1 \neq m_1\}\\
=& \Pr\{\hat{m}_2 \neq m_2 | \hat{\hat{m}}_1 = m_1 \} + \Pr\{ \hat{\hat{m}}_1 \neq m_1\}\\
=& \Pr\{e_2\} + \Pr\{e_1\}
\end{split}
\end{equation}
As shown by (\ref{eq:Pe}), an error occurs at the destination if either $\hat{\hat{m}}_1$ or $\hat{m}_2$ are not correctly decoded. In the sequel, we successively give upper bounds on $\Pr\{e_1\} \triangleq \Pr\{ \hat{\hat{m}}_1 \neq m_1\}$ and $\Pr\{e_2\} \triangleq \Pr\{\hat{m}_2 \neq m_2 | \hat{\hat{m}}_1 = m_1 \}$.

\subsection{Probability of error $\Pr\{e_1\}$ on $\hat{\hat{m}}_1$}\label{sec:Error1}

We first analyze the probability that the destination decoder makes an error while decoding the bin index $\hat{\hat{m}}_1$: \begin{equation}
\begin{split}
\Pr\{e_1\}  \triangleq &\Pr\{ \hat{\hat{m}}_1 \neq m_1\}\\
=& \Pr\{  \hat{\hat{m}}_1 \neq m_1 | \hat{m}_1 = m_1 \} \Pr\{ \hat{m}_1 = m_1\} \\
 & + \Pr\{  \hat{\hat{m}}_1 \neq m_1 | \hat{m}_1 \neq m_1 \} \Pr\{ \hat{m}_1 \neq m_1\}\\
\leq & \Pr\{  \hat{\hat{m}}_1 \neq m_1 | \hat{m}_1 = m_1 \} + \Pr\{ \hat{m}_1 \neq m_1\}\\
=& \Pr\{e_{12}\} + \Pr\{e_{11}\}
\end{split}
\end{equation}

An error on $\hat{\hat{m}}_1$ results either from an error at the relay who incorrectly decodes $\hat{m}_1$, which is denoted $\Pr\{e_{11}\}\triangleq \Pr\{ \hat{m}_1 \neq m_1\}$, or from an error at the destination given that the relay had correctly decoded the bin index $\hat{m}_1=m_1$, denoted $\Pr\{e_{12}\}\triangleq \Pr\{  \hat{\hat{m}}_1 \neq m_1 | \hat{m}_1 = m_1 \}$.

\smallskip

\noindent \textbf{Probability of error $ \Pr\{e_{11}\}$}

We first analyze the probability of error at the relay $\Pr\{e_{11}\}$. From the definition (\ref{eq:DecSetR}) of the relay decision set $\mathcal{S}_R$, it can be seen that two types of errors can happen at the relay decoder. The first type corresponds to the event $e_{11,m_1}=\{\bin_{m_1}\notin \mathcal{S}_R\}$. The second type of error occurs if there exists a $k\neq m_1$ such that $\bin_{k}\in \mathcal{S}_R$, event that we call $e_{11,k}$.
By the union bound, the probability of error $\Pr\{e_{11}\}$ is upper-bounded by
\begin{equation}\label{eq:Pe11UnionB}
\Pr\{e_{11}\} = \Pr\left\{ \bigcup_{k=0}^{M_R-1} e_{11,k} \right\} \leq \Pr\{e_{11,m_1}\} + M_R \Pr\{e_{11,k\neq m_1}\}.
\end{equation}
The probability of errors of type $1$ is upperbounded by
\begin{equation}\label{eq:Pe11m1}
\begin{split}
\Pr\{e_{11,m_1}\}
= & \Pr\{\forall l \in \bin_{m_1}, S_{RS,l} < A_R \}\\
= &\prod_{l\in \bin_{m_1}} \Pr\{S_{RS,l} < A_R \}\\
= & \Pr\{S_{RS,m} < A_R \} \Pr\{S_{RS,l \neq m} < A_R \}^{M_D-1}\\
\leq & \Pr\{S_{RS,m} < A_R \},
\end{split}
\end{equation}
where the second equality comes from the independence of variables $\{S_{RS,l}\}_l$.
From (\ref{eq:sigmak}), the sequence $\{|R_{RS,m}(n)|^2\}_n$ is a sequence of i.i.d. variables with mean $\sigma_{m}^2= 1+ \frac{a^2 P_S (T_s - 2 T_d)}{\theta N_0}$. By the weak law of large numbers $S_{RS,m}$ converges in probability to $\sigma_{m}^2$:
\begin{equation}\label{eq:WLLN}
\forall \nu >0, \lim_{N\rightarrow\infty}\Pr\left(\left|S_{RS,m}-\sigma_{m}^2\right|\geq\nu\right)=0
\end{equation}
The probability or error $\Pr\{S_{RS,m} < A_R \}$ can be written
\begin{equation}\label{eq:Pe11m1WLLN}
\begin{split}
\Pr\{S_{RS,m} < A_R \}
& = \Pr\left\{S_{RS,m} - \sigma_{m}^2 \leq - \epsilon \frac{a^2 P_S (T_s - 2 T_d)}{\theta N_0} \right\}\\
& \leq \Pr\left\{|S_{RS,m} - \sigma_{m}^2| \geq \epsilon \frac{a^2 P_S (T_s - 2 T_d)}{\theta N_0}  \right\},
\end{split}
\end{equation}
which decays to $0$ as $N$ grows large by (\ref{eq:WLLN}) for any $\epsilon>0$ and for any rate $R=R_1+R_2$. Errors of type $1$ are not rate-limiting, on the contrary to errors of type $2$ as we will now see.

The probability of errors of type $2$ is
\begin{equation}\label{eq:Pe11k}
\begin{split}
\Pr\{e_{11,k\neq m_1}\}
= & \Pr\{\exists l \in \bin_{k\neq m_1}, S_{RS,l} \geq A_R \}\\
= & \Pr\left\{\bigcup_{l \in \bin_{k\neq m_1}} \{ S_{RS,l} \geq A_R \} \right\}\\
= & \sum_{l \in \bin_{k\neq m_1}} \Pr \{ S_{RS,l} \geq A_R \} \\
= & M_D \Pr \{ S_{RS,l \neq m} \geq A_R \}
\end{split}
\end{equation}
where the two last equalities are due to the fact that for all $l\neq m$, variables $\{S_{RS,l}\}_l$ are i.i.d.
Using Markov's inequality gives the upper bound
\begin{equation}\label{eq:MarkovIneq}
\begin{split}
\Pr \{ S_{RS,l\neq m} \geq A_R \} &= \Pr\{ e^{r S_{RS,l\neq m}} \geq e^{rA_R}\} \\
& \leq \frac{\mathds{E}[e^{r S_{RS,l\neq m}}]}{e^{rA_R}} \mbox{ , } \forall r > 0
\end{split}
\end{equation}
$\{R_{RS,l\neq m}(n)\}_n$  is a set of i.i.d. $\mathcal{CN}(0,1)$ random variables. Therefore
\begin{equation}\label{eq:meanexp}
\begin{split}
\mathds{E}[e^{rS_{RS,l\neq m}}]& = \prod_{n=1}^N \mathds{E}[e^{\frac{r}{N}|R_{RS,l\neq m}(n)|^2}] \\\
 & = \left(1-\frac{r}{N}\right)^{-N} \mbox{ , } \forall \frac{r}{N} \in ]0,1[,
\end{split}
\end{equation}
recalling that $|R_{RS,l\neq m}(n)|^2$ is exponentially distributed with parameter $1$ and has a moment generating function $M(t)=\E[e^{t |R_{RS,l\neq m}(n)|^2}]=1/(1-t)$ for all $t\in[0,1[$.
Using (\ref{eq:MarkovIneq}) and (\ref{eq:meanexp}) in (\ref{eq:Pe11k}) yields
\begin{equation}\label{eq:Pe11kbound}
\Pr\{e_{11,k\neq m_1}\} \leq M_D e^{-N g_{A_R}(r/N)} \mbox{ , }\forall \frac{r}{N} \in ]0,1[
\end{equation}
where
\begin{equation}
g_{A_R}(u)\triangleq A_R u + \ln (1-u).
\end{equation}
The maximum of $g_{A_R}$ is reached for $u_{\max}=1-\frac{1}{A_R} \in ]0,1[$
\begin{equation}\label{eq:gAmax}
g_{A_R}(1-1/A_R)= A_R -1 - \ln (A_R).
\end{equation}
Using (\ref{eq:gAmax}) in (\ref{eq:Pe11kbound}) gives
\begin{equation}\label{eq:Pe11kbound2}
\Pr\{e_{11,k\neq m_1}\} \leq M_D e^{-N (A_R -1 - \ln A_R )}.
\end{equation}
Finally, using (\ref{eq:Pe11kbound2}) and (\ref{eq:Pe11m1}), we can rewrite the upper bound on $\Pr\{e_{11}\}$ (\ref{eq:Pe11UnionB})
\begin{equation}\label{eq:Pe11bound}
\begin{split}
\Pr\{e_{11}\}
& \leq \Pr\{S_{RS,m} < A_R \} + M_R M_D e^{-N (A_R -1 - \ln A_R )}\\
& = \Pr\{S_{RS,m} < A_R \} + e^{-N ( - \frac{RT_s}{\theta} + A_R -1 - \ln A_R )}.
\end{split}
\end{equation}
From (\ref{eq:Pe11m1WLLN}) the first term in (\ref{eq:Pe11bound}) vanishes to $0$ as $N$ grows large for any rate $R$, whereas the second term vanishes to $0$ if
\begin{equation}\label{eq:RboundPs}
\begin{split}
R <& \:\: \frac{\theta}{T_s} (A_R -1 - \ln A_R)\\
=& \:\:  (1-\epsilon)\frac{a^2 P_S}{N_0}\left(1-2\frac{T_d}{T_s}\right)\\
&\:\:  - \frac{\theta}{T_s} \ln \left(1+ (1-\epsilon) \frac{a^2 P_S (T_s - 2 T_d)}{\theta N_0}\right)\\
 \xrightarrow[\theta \rightarrow 0]{}& \:\:  (1-\epsilon)\frac{a^2 P_S}{N_0}\left(1-2\frac{T_d}{T_s}\right)
\end{split}
\end{equation}
Choosing $\epsilon$ arbitrarily close to zero, and $T_s$ arbitrarily close to $T_c$, (\ref{eq:RboundPs}) becomes
\begin{equation}\label{eq:Rbound}
R < \frac{a^2 P_S}{N_0}\left(1-2\frac{T_d}{T_c}\right).
\end{equation}
This bound is rather constraining since it means that message $m$ can be reliably decoded at the relay, although the relay actually forwards the bin index $\hat{m}_1$ only.

\smallskip

\noindent \textbf{Probability of error $ \Pr\{e_{12}\}$}

We now analyze the probability of error $\Pr\{e_{12}\}$ on $\hat{\hat{m}}_1$ at the destination decoder. From the definition (\ref{eq:DecSet1}) of decision set $\mathcal{S}_1$, two types of errors can happen at the destination decoder. The first type corresponds to the event $e_{12,m_1}=\{m_1 \notin \mathcal{S}_{1}| \hat{m}_1 = m_1 \}$, and the second type to $e_{12,k}=\{ k \neq m_1, k \in \mathcal{S}_{1}  | \hat{m}_1 = m_1 \}$.
By the union bound, the probability of error $\Pr\{e_{12}\}$ is upper-bounded by
\begin{equation}\label{eq:Pe12UnionB}
\begin{split}
\Pr\{e_{12}\}
\leq & \Pr\{e_{12,m_1}\} + M_R \Pr\{e_{12,k\neq m_1}\}\\
= & \Pr \{ S_{DR,\hat{m}_1} < B_R | \hat{m}_1 = m_1\} \\
  & + M_R \Pr \{ S_{DR,k\neq\hat{m}_1} \geq B_R | \hat{m}_1 = m_1\}
\end{split}
\end{equation}
The second term in (\ref{eq:Pe12UnionB}) can be upper bounded using Markov's inequality as in (\ref{eq:MarkovIneq}) and (\ref{eq:meanexp})
\begin{equation}\label{eq:MarkovIneq12}
\Pr \{ S_{DR,k\neq\hat{m}_1} \geq B_R | \hat{m}_1 = m_1\} \leq e^{-N g_{B_R}(r/N)} \mbox{ , }\forall \frac{r}{N} \in ]0,1[
\end{equation}
where
\begin{equation}
g_{B_R}(u)\triangleq B_R u + \ln (1-u).
\end{equation}
Using the maximum $g_{B_R}(1-1/B_R)=B_R -1 - \ln B_R$ and (\ref{eq:MarkovIneq12}) in (\ref{eq:Pe12UnionB}) gives
\begin{equation}\label{eq:Pe12bound}
\begin{split}
& \Pr\{e_{12}\} \\
& \leq \Pr \{ S_{DR,\hat{m}_1} < B_R | \hat{m}_1 = m_1\} + M_R e^{-N (B_R -1 - \ln B_R)}\\
& =  \Pr \{ S_{DR,\hat{m}_1} < B_R | \hat{m}_1 = m_1\} + e^{-N (- \frac{R_1T_s}{\theta} + B_R -1 - \ln B_R)}
\end{split}
\end{equation}
As in (\ref{eq:WLLN}) and (\ref{eq:Pe11m1WLLN}) the weak law of large numbers ensures that $S_{DR,\hat{m}_1}$ converges in probability to $\sigma_{R,\hat{m}_1}^2= 1+ \frac{b^2 P_R (T_s - 2 T_d)}{\theta N_0}$, and thus that the first term $\Pr \{ S_{DR,\hat{m}_1} < B_R | \hat{m}_1 = m_1\}$ in (\ref{eq:Pe12bound}) decays to $0$ as $N$ grows large for any $ \epsilon_1 >0$ and any rate $R=R_1+R_2$. Using the same reasoning as in (\ref{eq:RboundPs}), and choosing $\epsilon_1$ arbitrarily close to zero, it can be shown that the second term in (\ref{eq:Pe12bound}) will vanish to $0$ when $N \rightarrow + \infty$, and $\theta \rightarrow 0$ if
\begin{equation}\label{eq:R1bound}
R_1 < \frac{\gamma b^2 P_S}{N_0}\left(1-2\frac{T_d}{T_c}\right).
\end{equation}

To summarize, in this subsection, the probability of error on $\hat{\hat{m}}_1$, $ \Pr \{e_1\} \leq \Pr\{e_{12}\} + \Pr\{e_{11}\}$ has been shown to decay to $0$ if $ R < \frac{a^2 P_S}{N_0}\left(1-2\frac{T_d}{T_c}\right)$ and $R_1 < \frac{b^2 P_S}{N_0}\left(1-2\frac{T_d}{T_c}\right) $.

\subsection{Probability of error $\Pr\{e_2\}$ on $\hat{m}_2$}\label{sec:Error2}

In this subsection, the probability that the destination decoder makes an error when decoding $\hat{m}_2$ is analyzed, assuming that the bin index $\hat{\hat{m}}_1$ was correctly decoded: $\Pr\{e_2\} \triangleq \Pr\{\hat{m}_2 \neq m_2 | \hat{\hat{m}}_1 = m_1 \}$. Given $\hat{\hat{m}}_1=m_1$, decoding $\hat{m}_2$  is equivalent to identifying the frequency $m$ in $\bin_{\hat{\hat{m}}_1}$, which contains $M_D$ frequencies. The upper bound on $\Pr\{e_2\}$ is obtained by following the same reasoning as for $\Pr\{e_{12}\}$ in Section \ref{sec:Error1}. From the definition (\ref{eq:DecSet2}) of decision set $\mathcal{S}_2$, two types of error can occur at the decoder, and the union bound gives
\begin{equation}\label{eq:Pe2UnionB}
\begin{split}
&\Pr\{e_{2}\}\\
& \leq \Pr \{ S_{DS,m} < B_D | \hat{\hat{m}}_1 = m_1\} \\
  & + M_D \Pr \{ S_{DR,k\neq m} \geq B_D | \hat{\hat{m}}_1 = m_1\}\\
&\leq  \Pr \{ S_{DS,m} < B_D | \hat{\hat{m}}_1 = m_1\} + M_D e^{-N (B_D -1 - \ln B_D)}\\
& = \Pr \{ S_{DS,m} < B_D | \hat{\hat{m}}_1 = m_1\} + e^{-N (- \frac{R_2T_s}{\theta} + B_D -1 - \ln B_D)},
\end{split}
\end{equation}
where the second inequality is due to Markov's inequality. The weak law of large numbers ensures that the first term in (\ref{eq:Pe2UnionB}) decays to $0$ as $N$ grows large for any $\epsilon_2>0$, and any rate $R=R_1+R_2$, while the second term in (\ref{eq:Pe2UnionB}) vanishes to $0$ when $N \rightarrow + \infty$, and $\theta \rightarrow 0$ if
\begin{equation}\label{eq:R2bound}
R_2 < \frac{P_S}{N_0}\left(1-2\frac{T_d}{T_c}\right).
\end{equation}

The case $a^2>1$ can be summarized by combining the results (\ref{eq:Rbound}), (\ref{eq:R1bound}), and (\ref{eq:R2bound}) on $\Pr\{e_1\}$, and $\Pr\{e_2\}$: the total probability of errors $P_e \leq \Pr\{e_1\} + \Pr\{e_2\}$ vanishes to $0$ for all rates $R=R_1+R_2$ such that
\begin{equation}
R \leq \min \{a^2, 1+\gamma b^2\} \frac{P_s}{N_0} \left(1-2\frac{T_d}{T_c}\right),
\end{equation}
which is equal to (\ref{eq:LBfading}) when $a^2>1$. This concludes the proof of 1) in \emph{Theorem \ref{th:AchRateFading}}.

\vspace{-0.12cm}

% references section

\bibliographystyle{./Biblio/IEEEtran}

\bibliography{./Biblio/IEEEabrv,./Biblio/bibLowSNR,./Biblio/bibLargeNet}

\end{document}

%% file: sty/macros.tex
\newtheorem{theorem}{Theorem}%[section]
\newfont{\mbb}{msbm10 scaled 1100}

\definecolor{red}{RGB}{210,20,50}
\newcommand{\red}{\textcolor{red}}
\definecolor{lblue}{rgb}{.15,.4,.8}

\definecolor{green}{RGB}{75,180,130}
\newcommand{\green}{\textcolor{green}}
\definecolor{blue}{RGB}{0,0,255}
\newcommand{\blue}{\textcolor{blue}}

\definecolor{orange}{RGB}{255,128,0}

\definecolor{dgreen}{RGB}{0,151,0}

\def\E{{\rm E}} % expectation
\def\Pr{{\rm Pr}} % probability
\def\bin{{\rm bin}}
\def\ind{{\mathds{1}}}% indicator function
\def\N{{\mathds{N}}} % set of non-negative integers
\def\R{{\mathds{R}}} % set of real number
\def\Rplus{{\mathds{R}^+}} % set of non-negative real numbers
\def\b0{{\bf 0}}